\documentclass[twocolumn,amsmath,amssymb,prl,aps,showkeys]{revtex4-1}
\usepackage{graphicx}
\usepackage{amsmath,amssymb}
\usepackage{mathrsfs}
\usepackage{color}
\usepackage{afterpage}
\usepackage[normal]{subfigure}
\usepackage{natbib}

\begin{document}
\title{Theoretical evidence for unexpected O-rich phases at corners of MgO surfaces}
\author{Saswata Bhattacharya$^1$, Daniel Berger$^2$, Karsten Reuter$^2$, Luca M. Ghiringhelli$^3$, Sergey V. Levchenko$^3$} 
\affiliation{$^1$ Indian Institute of Technology Delhi, New Delhi 110016, India\\ $^2$Chair for Theoretical Chemistry and Catalysis Research Center, Technical University Munich, Lichtenbergstr. 4, D-85747 Garching, Germany, $^3$Fritz-Haber-Institut der Max-Planck-Gesellschaft, Faradayweg 4-6, D-14195 Berlin, Germany}
\date{\today}
\begin{abstract}

Realistic oxide materials are often semiconductors, in particular at elevated temperatures, and their surfaces contain undercoordiated atoms at structural defects such as steps and corners.  Using hybrid density-functional theory and {\em ab initio} atomistic thermodynamics, we investigate the interplay of bond-making, bond-breaking, and charge-carrier trapping at the corner defects at the (100) surface of a $p$-doped MgO in thermodynamic equilibrium with an O$_2$ atmosphere. We show that by manipulating the coordination of surface atoms one can drastically change and even reverse the order of stability of reduced versus oxidized surface sites.
\end{abstract}
\maketitle

In surface catalysis, point defects often play the role of active sites. It is widely accepted that O vacancies are important active sites at metal-oxide surfaces~\cite{Zhen2005, Balint2001}. This implies that the concentration of the vacancies at reaction conditions is sufficiently high (above few parts per million). Typically, defect properties are studied under ultra-high vacuum conditions, when the concentration is higher due to low oxygen partial pressure $p_{\rm O_2}$. It's therefore important to explore how many vacancies remain stable under the O-rich conditions of technological catalysis.

Impurities and intrinsic defects often convert materials that are ideally band-gap insulators (e.g., TiO$_2$, ZnO, MgO) into semiconductors. We have recently demonstrated that this charge-carrier doping (either accidental or intentional) can strongly influence the material's surface chemistry.~\cite{Richter2013} The formation energy of an isolated oxygen vacancy at the (100) surface of MgO is reduced by about 6~eV when the charge carriers (holes) are available in the material. As a result, even at low doping levels ($<10^{17}$~cm$^{-3}$) the concentration of the charged O vacancies at a MgO (100) terrace at $T$ = 400~K and $p_{\rm O_2}$ = 1~atm exceeds 0.01~at.\%. As discussed below, we find that even at these O-rich conditions the concentration of O-ad-species at the $p$-doped MgO surface remains vanishingly small.

Realistic oxide surfaces, however, are rarely atomically flat. They contain structural defects such as steps and corners, where atoms have a coordination different from the flat surface (terrace). As one may expect, the formation energy of a neutral O vacancy at steps and corners at the MgO (100) surface is lower than at the terrace~\cite{Sushko2000153,Sicolo2013}. 
However, small MgO clusters in an O$_2$ atmosphere contain an {\em excess} rather than deficiency of oxygen compared to the stoichiometric oxide, even at high temperatures~\cite{Bhattacharya2013}. Thus, it is not clear {\em a priori} how the lower coordination will influence binding of an O atom and the charge-carrier trapping energy at a doped oxide surface.

In this Letter, we address the above problem. Our material of choice is MgO -- an important catalytic material and a widely studied prototypical system. We focus on the corners at the $p$-doped~\footnote{A semiconducting $p$-doped MgO can be produced by, e.g., doping with Li~\cite{Chen1980,tardio2002p}. Even without intentional doping, nominally pure MgO becomes a $p$-type semiconductor at temperatures above 800~K~\cite{Freund1993}, which is comparable or even below the typical temperatures for some of the catalytic applications of MgO.} MgO (100) surface, since the differences with the terrace are expected to be more pronounced in this case. The step edges will be addressed elsewhere. MgO surfaces with a high concentration of corner sites can be prepared by annealing MgO (111) surfaces or by producing nanocrystalline MgO. We calculate relative concentrations of corner defects of different types (vide infra, the O-vacancy, as well as O$_1$/O$_2$-ad-species), using density-functional theory (DFT) and {\em ab initio} atomistic thermodynamics~\cite{Scheffler1986}. The doping is considered only as means of fixing the chemical potential of the electrons ($\mu_e$, see below).

\begin{figure}[t!]
\includegraphics[width=0.8\columnwidth,clip]{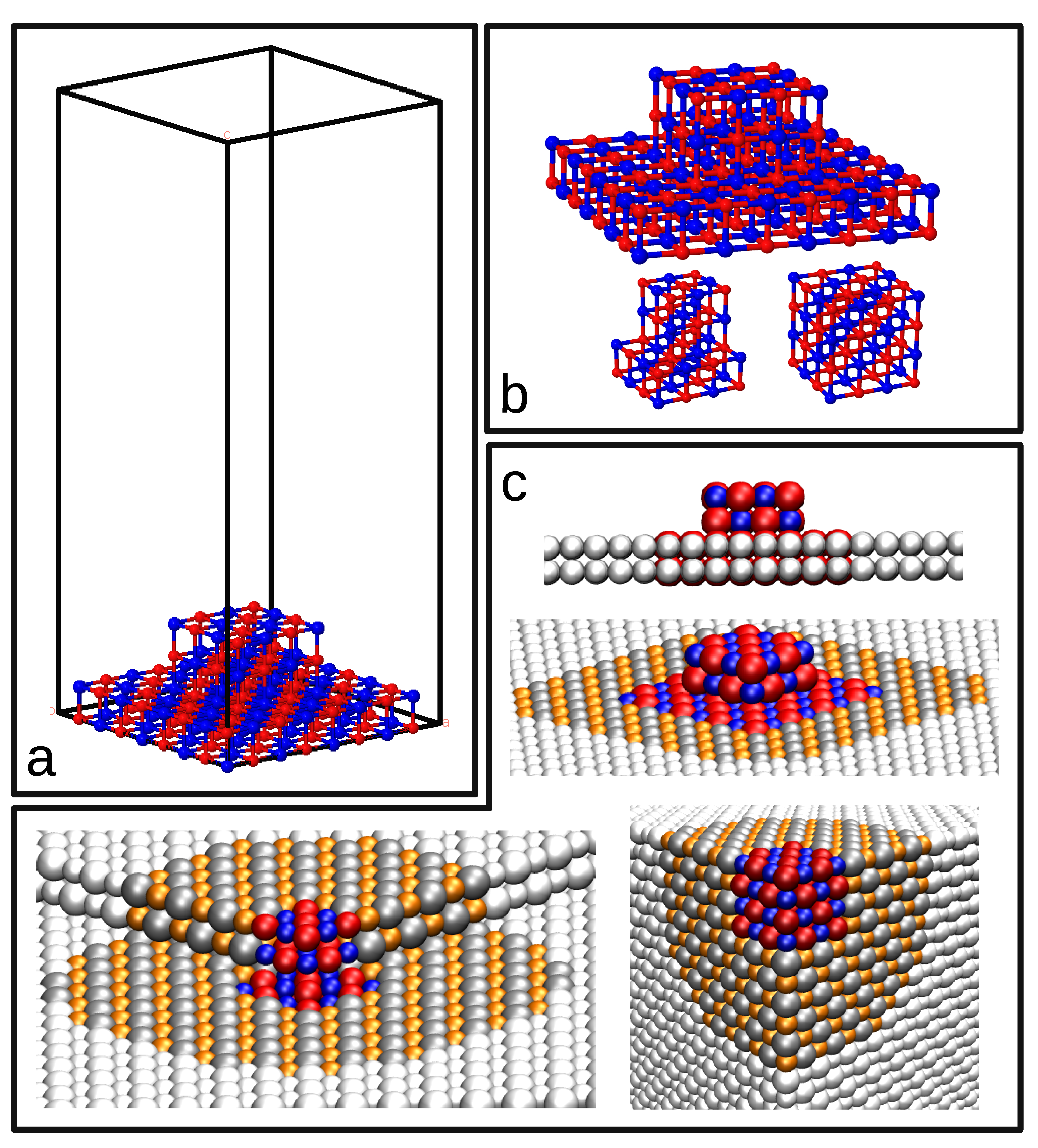}
\caption{(color online). (a) Periodic model of Mg$_{80}$O$_{80}$ corner. (b) Mg$_{80}$O$_{80}$, Mg$_{25}$O$_{25}$ and Mg$_{32}$O$_{32}$ clusters, and (c) the corresponding embedded cluster models for Mg$_{25}$O$_{25}$ and Mg$_{32}$O$_{32}$ (Mg$_{80}$O$_{80}$ embedded model is not shown). The red (blue) spheres represent O (Mg) atoms in the QM region; white spheres are point charges ($+2$ for Mg$^{2+}$ and $-2$ for O$^{2-}$) fixed to the perfect lattice positions; gray (yellow) spheres are the polarizable $-2$ ($+2$) charges (see text).}
\label{fig1}
\end{figure} 
Corner defects are modelled by embedded clusters (see Fig.~\ref{fig1}). Corners separated from the underlying surface by one, two, or infinite number of layers are considered. 
We use a QM/MM-type embedding~\cite{Bernstein2009}, where an atomic cluster treated at the {\em ab initio} level (the quantum-mechanical part, QM) is embedded into a region treated at the level of empirical interatomic potentials (the molecular-mechanics part, MM). The MM region is divided into an inner ``active'' part (polarizable), where ions are allowed to relax their positions, and an outer part where ions are constrained to their lattice positions. Direct linkage of the field of MM point charges with the QM region results in a spurious over-polarization of the wavefunction (charge leakage) towards the positive point charges (MM cations). Therefore, norm-conversing non-local pseudopotentials~\cite{KleinmanBylander1982} are used instead of point charges for the MM cations adjacent to the QM part. The {\em ab initio} calculations are performed with the {\tt FHI-aims} program package~\cite{Blum2009}, where the QM/MM embedding infrastructure has been recently implemented~\cite{Berger2014, Berger2015}.
The O sites in the active MM region are represented as a core (O$_{c}$) and shell (O$_{s}$) site, interacting via a harmonic potential. Mg sites in the active MM region interact via a Buckingham-type potential with O$_{c}$ and O$_{s}$ sites.  
We fit the interatomic-potential parameters for the MM region to match the lattice parameters and dielectric constant of the QM region 
(with the functional applied) (see SI).
The total energy of the system is defined as: $E^{\rm tot} = E^{\rm QM} + E^{\rm MM}$ and is minimized iteratively in the {\tt ChemShell} environment \cite{Sherwood2003,Sokol2004}, connecting {\tt FHI-aims} with the MM driver {\tt GULP}\cite{GULP}. 
In a given geometry {\tt FHI-aims} evaluates $E^{\rm QM}$ and corresponding forces acting on all particles (QM atoms, embedding point charges, and pseudopotentials). The MM driver evaluates further forces on embedding point charges at the level of interatomic potentials, and together with the QM forces performs geometry optimization.
Full polarization response is calculated by relaxation of all QM atoms, pseudopotentials and MM particles within the active MM region. Long-range contributions beyond 
the active region are corrected for analytically (see SI).


The QM region is treated using DFT. The full atomic relaxation is done with PBE~\cite{Perdew1996, Perdew1997}, and total energy calculations are performed at these geometries with the hybrid functional HSE06~\cite{Heyd2006}. First-principles dispersion correction is included via the Tkatchenko-Scheffler scheme~\cite{Tkatchenko2009}. The high-accuracy basis sets and numerical grids are defined by the tight settings~\cite{Blum2009} that have been used throughout our calculations.

For an accurate description of defect charging, the embedding scheme should closely reproduce the electronic structure of the corresponding extended system, in particular the position of the defect levels with respect to (w.r.t.) the bulk Fermi level. We check this by comparing the electronic structure of the embedded cluster model of a pristine O-terminated corner with a periodic slab model (see SI).  Since we focus on $p$-doped MgO, the Fermi level is set to the valence band maximum (VBM). The position of the highest occupied state (HOMO) for the corner defect w.r.t. the bulk Fermi level is determined as follows. First, the position of the Fermi level w.r.t. the vacuum level is calculated using the HSE06 functional. This is done by calculating the position of the Mg 1s core state in the third layer of a 5-layer MgO (100) slab w.r.t. vacuum. By aligning the positions of the Mg 1s core states in the slab and in the bulk, the position of the VBM w.r.t. vacuum is then calculated. The result is -7.5~eV. Next, the position of the HOMO w.r.t. vacuum is calculated from the embedded cluster model. To ensure that no artificial shift of the HOMO relative to the vacuum level is induced by the embedding, we compare the positions of the HOMO for both the embedded cluster and the periodic slab models of the pristine neutral corner, calculated with the PBE~\cite{Perdew1996, Perdew1997} functional. The periodic model was converged in terms of the number of layers and lateral dimensions. The resulting difference is found to be about 0.2~eV. We conclude that we have an uncertainty of at most $\pm 0.2$ eV (depending on the type of corner) in the position of the defect levels w.r.t. the Fermi level due to embedding.

In order to identify the relevant defect structures for the QM region, we have studied the relative stability of a set of parallelepipedal  [Mg$_N$O$_{(x+N)}$)]$^q$ clusters of various sizes [$N$ = 18, 24, 32, 108] with a defect at one of the corners [$x$ = -1 (O-vacancy), 0 (pristine), 1 (O$_1$-ad-species), 2 (O$_2$-ad-species)]. The charge state $q$ is varied from -2 to +2 [$q$ = -2, -1, 0, 1, 2]. By using our cascade genetic algorithm [cGA]~\cite{Bhattacharya2013, Bhattacharya2014}, we found that for all defects global minimum structures have the defect located at the O-terminated corner and not at the edges or faces of the parallelepipedal structures (Fig. \ref{fig3}, lower panel). The greater stability of the oxygen ad-species at the O- versus Mg-terminated corner is explained by the stronger basicity of the undercoordinated O atom that can donate electrons to the additional electrophilic O species.

We analyze the thermodynamic stability of the defected clusters w.r.t. the pristine clusters by means of {\em ab initio} atomistic thermodynamics. We consider a reservoir of gas-phase O$_2$ molecules, characterized by the chemical potential $\mu_{\textrm{O}_2}$. For clusters [Mg$_N$O$_{(N+x)}$]$^{q}$ with charge $q$, the formation Gibbs free energy $[\Delta G^{q}(T, p_{\textrm{O}_2})]$ w.r.t. the host (neutral stoichiometric cluster) is
\begin{equation}
\label{eqn5}
\Delta G^{q}(T, p_{\textrm{O}_2}) = \Delta E_{\rm f} - x\Delta \mu_\textrm{O} (T, p_{\textrm{O}_2}) + q\mu_\textrm{e}\,
 \end{equation}
where $\Delta E_{\rm f}$ = $E_\textrm{defect} ^{q} - E_\textrm{host}^{q=0}-\frac{x}{2}E_{\rm O_2}$ is the DFT defect formation energy, $\Delta \mu_\textrm{O} = \frac{1}{2} (\mu_{\textrm{O}_2}-E_{\rm O_2})$, and $E_{\rm O_2}$, $E_\textrm{defect} ^{q}$, $E_\textrm{host}^{q=0}$ are the total energies of the oxygen molecule, the cluster with defect, and the cluster without defect, respectively. Vibrational contributions to the free energies of formation are found to be around 0.1~meV for the systems studied here, and are therefore neglected.

We find that at realistic temperatures and pressures (e.g., $T$ = 300~K, $p_{\rm O_2} \sim 1$ atm) O$_1$- or O$_2$-ad-species at the oxygen corners of pristine [(MgO)$_N$]$^q$ clusters are favored for all studied charge states, over a wide range of electronic chemical potential $\mu_e$ (phase diagrams are given in the SI). In fact, the O$_1$/O$_2$-ad-species are stable even at some values of $\mu_e$, for which O-vacancies are preferred at flat (100) surfaces of $p$-doped MgO. 

Following the findings for the isolated parallelepipedal clusters, we have created several embedded cluster models,  taking as a starting point for the QM region Mg$_{32}$O$_{32+x}$ (Fig.~\ref{fig1}b, bottom right cubic structure), Mg$_{25}$O$_{25+x}$ (Fig.~\ref{fig1}b, bottom left step structure), and  Mg$_{20}$O$_{21+x}$ (not shown). Here, $x=0$ represents the pristine corners, while $x=$ -1, 1 and 2 are respectively O-vacancy, O$_1$-ad-species and O$_2$-ad-species. These clusters are embedded in the MM-regions as explained above (see Fig.~\ref{fig1}c). 
The initial geometries for the embedded clusters are created by cutting out the defected corner (for $x=$ -1, 1 and 2) and its nearest neighbors from the isolated parallelepipedal clusters and replacing it with the corresponding part at the corner of interest of the otherwise perfect embedded cluster.
The number of relaxed atoms in the QM region were expanded until the convergence of the formation energies of the defects is achieved. Similarly, convergence of electrostatic potential w.r.t. the number of point charges is also thoroughly tested (see SI). 

\begin{figure}[t!]
\includegraphics[width=0.80\columnwidth,clip]{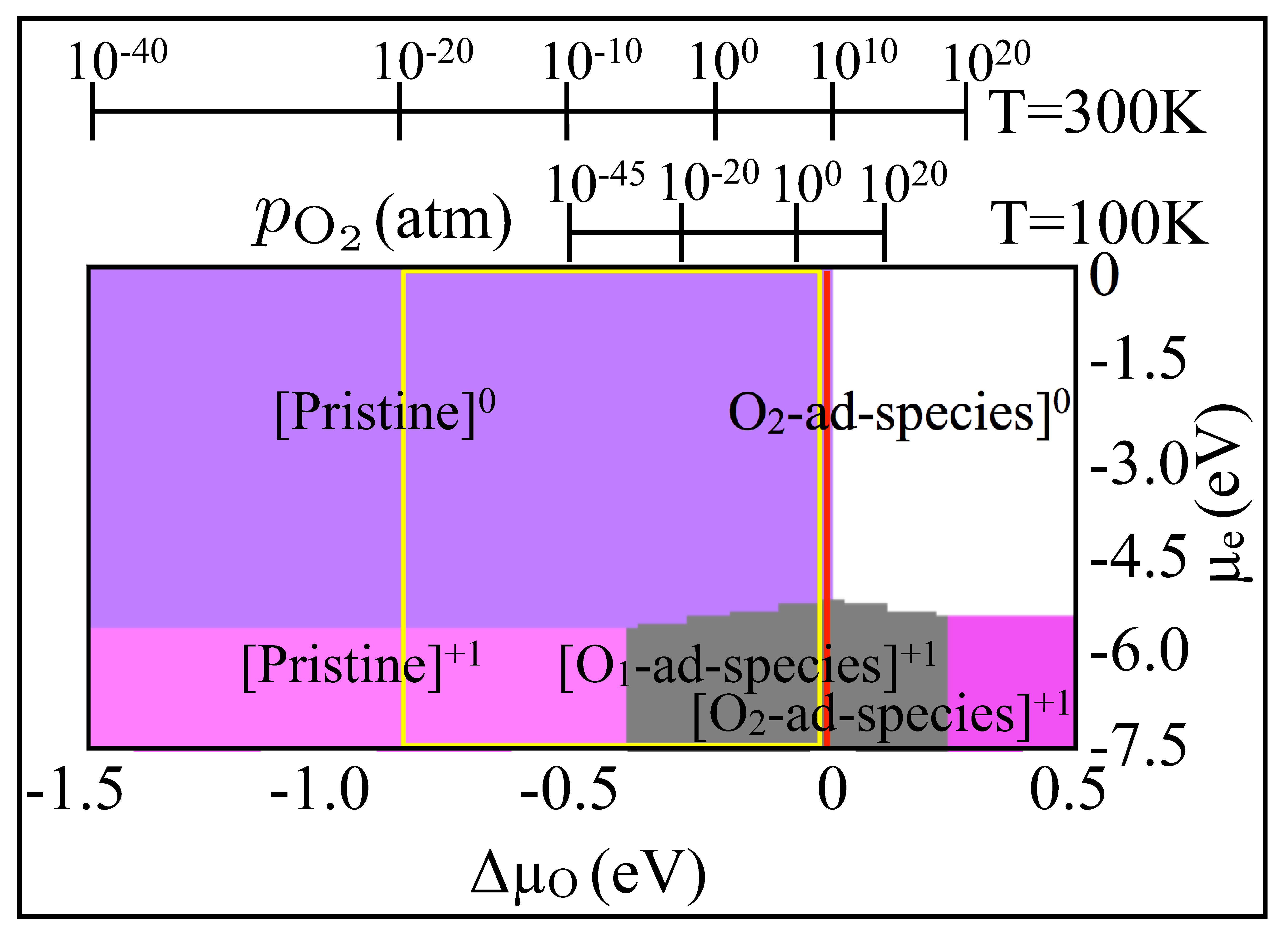}
\caption{Phase diagram for defects at corners of MgO (100) in an oxygen atmosphere, calculated using embedded [Mg$_{32}$O$_{32+x}$]$^q$ clusters, with $x=-1$ (vacancy), 0 (pristine), 1 (O$_1$-ad-species), and 2 (O$_2$-ad species), and $q=$ -2, 1, 0, 1, and 2. The energy zero on the $\mu_e$ axis~\footnote{This $\mu_e$ includes the shift associated with the band bending due to the formation of O vacancies at terraces (see text).} corresponds to the vacuum level. The yellow rectangle represents the region relevant for catalytic applications, while the red vertical line represents the O-rich limit.
}
\label{fig2}
\end{figure}

\begin{figure}[b!]
\includegraphics[width=0.80\columnwidth,clip]{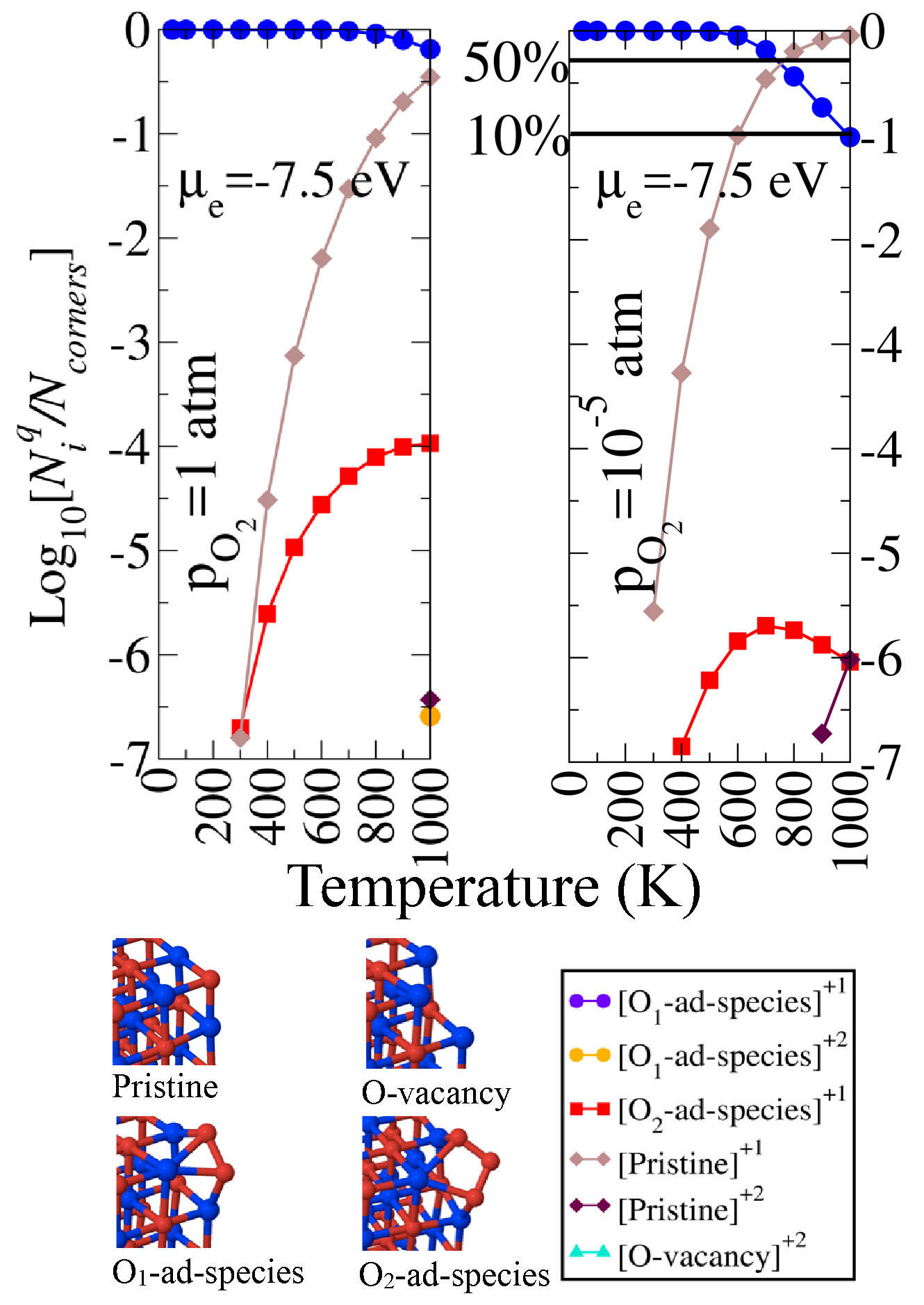}
\caption{(color online). Concentration of defects at corners at $p_{\textrm{O}_2}$ = 1~atm (left) and $p_{\textrm{O}_2}$ = 10$^{-5}$~atm (right) for $p$-doped MgO ($\mu_e = -7.5$~eV). The calculated concentrations include the effects of band bending for a dopant concentration N$_\textrm{D}$=10$^{16}$ cm$^{-3}$. There is a negligible change in the defect concentrations when the dopant concentration is increased to 10$^{22}$ cm$^{-3}$.}
\label{fig3}
\end{figure}

The thermodynamic stability of the corner defects at an extended (100) surface of $p$-doped MgO was then studied using the embedded cluster models and Eq.~\ref{eqn5}. $E_\textrm{defect}^q$ now refers to the embedded cluster with $x$ additional (removed when $x=$ -1) oxygen atoms and a charge $q$; while the host is the stoichiometric neutral embedded cluster. Here we show results for the defects at the corners formed by O-terminated multi-layer (three or more layers) step edges. The results for the corners formed by two-layer and monolayer step edges are similar (see SI). The calculated phase diagram is shown in Fig.~\ref{fig2}.  Clearly, O$_1$/O$_2$-ad-species are stabilized also at the corners at the extended surface. Note that the extra stability of O$_1$/O$_2$-ad-species over O-vacancies is preserved over a wide range of $\mu_e$ (see SI).

We also calculate the concentration of the defects as function of the thermodynamic parameters. In the limit of a small concentration of the corners, when the interaction between them can be neglected, the defects will obey the Fermi-Dirac statistics (as identical particles that cannot occupy the same site more than once). As has been shown previously~\cite{Richter2013}, formation of O vacancies at the (001) surface of $p$-doped MgO at realistic ($T$, $p_{\rm O_2}$) conditions and doping concentrations $N_{\rm D}$ results in the formation of a space-charge layer under the surface and concomitant band bending. For a small concentration of the corner sites, contribution of defects at these sites to the band bending can be neglected. In this case, the effect of the band bending on the formation energies of defects at corners is simply taken into account by shifting $\mu_e$ in Eq.~\ref{eqn5} upwards by the value of the band bending (since the band bending shifts the defect electronic levels down with respect to the bulk Fermi level), which is a function of $T$, $p_{\rm O_2}$, and $N_{\rm D}$~\cite{Richter2013}.

If $N$ is the total number of corners and $\Delta G_n$ is the formation Gibbs free energy relative to the pristine neutral corner of a given type-$n$ defect, then the number of corners with type-$n$ defect, $N_n$, is 
\begin{equation}
N_n = (N - \Sigma_{m\neq n} N_m) \frac{1}{\exp (\beta \Delta G_n) + 1}
\label{eqn7}
\end{equation}
where and $\beta = 1/k_\textrm{B} T$ and $N_m$ is any other considered defect type. A similar equation is written for each considered defect type.
The solution of the set of coupled equations is: 
\begin{equation} 
\nonumber \frac{N_n}{N} = \frac{\exp (-\beta \Delta G_n)}{1+\Sigma_m \exp (- \beta \Delta G_m)}\,
\label{eqn_conc}
\end{equation}

Fig.~\ref{fig3} shows the concentration of different kinds of defects at varying temperature and constant pressure. At $\mu_e$=$-7.5$ eV and $p_{\textrm{O}_2}$=1 atm, O$_1$-ad-species with a trapped hole ([O$_1$-ad-species]$^{+1}$ in Fig.~\ref{fig3}) at the corners of MgO surface are the most abundant species at all temperatures. Only at temperatures around 900~K and above another defect starts to compete with it, namely the pristine corner with a trapped hole ([Pristine]$^{+1}$ in Fig.~\ref{fig3}). At the high temperature, about 0.01\% of all O-terminated corners should have an O$_2$ molecule adsorbed on them, again with a trapped single hole ([O$_2$-ad-species]$^{+1}$ in Fig.~\ref{fig3}). Interestingly, surface O$_3^-$ species have been previously proposed to explain transmission electron diffraction measurements of reconstructed MgO (111) surfaces, which are expected to expose a high coverage of corner sites.~\cite{Richard_PRL1998} In fact, metal-superoxo and metal-ozonide complexes and their interconversion play an important role in organometallic chemistry and biochemistry, where they are well known as $\eta$ side groups. In our case, however, the fine-tuning of the chemical reactivity of metal centers is achieved by controlling the concentration of charge carriers and coordination at the inorganic surface sites, instead of organic ligands.

The estimated error of $\pm$0.2~eV in the position of the defect levels w.r.t. the electronic chemical potential ($\mu_e$) has no qualitative effect on the calculated relative concentrations (see SI).  
At lower O$_2$ pressures, pristine corner with a trapped hole ([Pristine]$^{+1}$ in Fig.~\ref{fig3}) starts to compete with [O$_1$-ad-species]$^{+1}$ at lower temperatures, but these two defects remain dominant in the considered temperature range. Another interesting results is that the concentration of oxygen vacancies with two trapped holes ([O-vacancy]$^{+2}$) at corners is found to be negligible at realistic conditions, which is in drastic contrast to the terrace sites. 

We have checked the stability of an O atom adsorbed at the (001) terrace of $p$-doped MgO in charge states 0, +1, and +2. According to our HSE06 results, the adsorbed O atom is at least 1.8~eV less stable (depending on the charge state) than the doubly charged O vacancy in the dilute limit (and by 1.3~eV for one defect per 2$\times$2 surface unit cell) even under most O-rich conditions ($\mu_\textrm{O}$ = 0). 
Thus, although at corners O ad-atoms are abundant defects, at the terraces their concentration is very small, and their effect on the band bending can be neglected.

\begin{figure}[t!]
\includegraphics[width=0.80\columnwidth,clip]{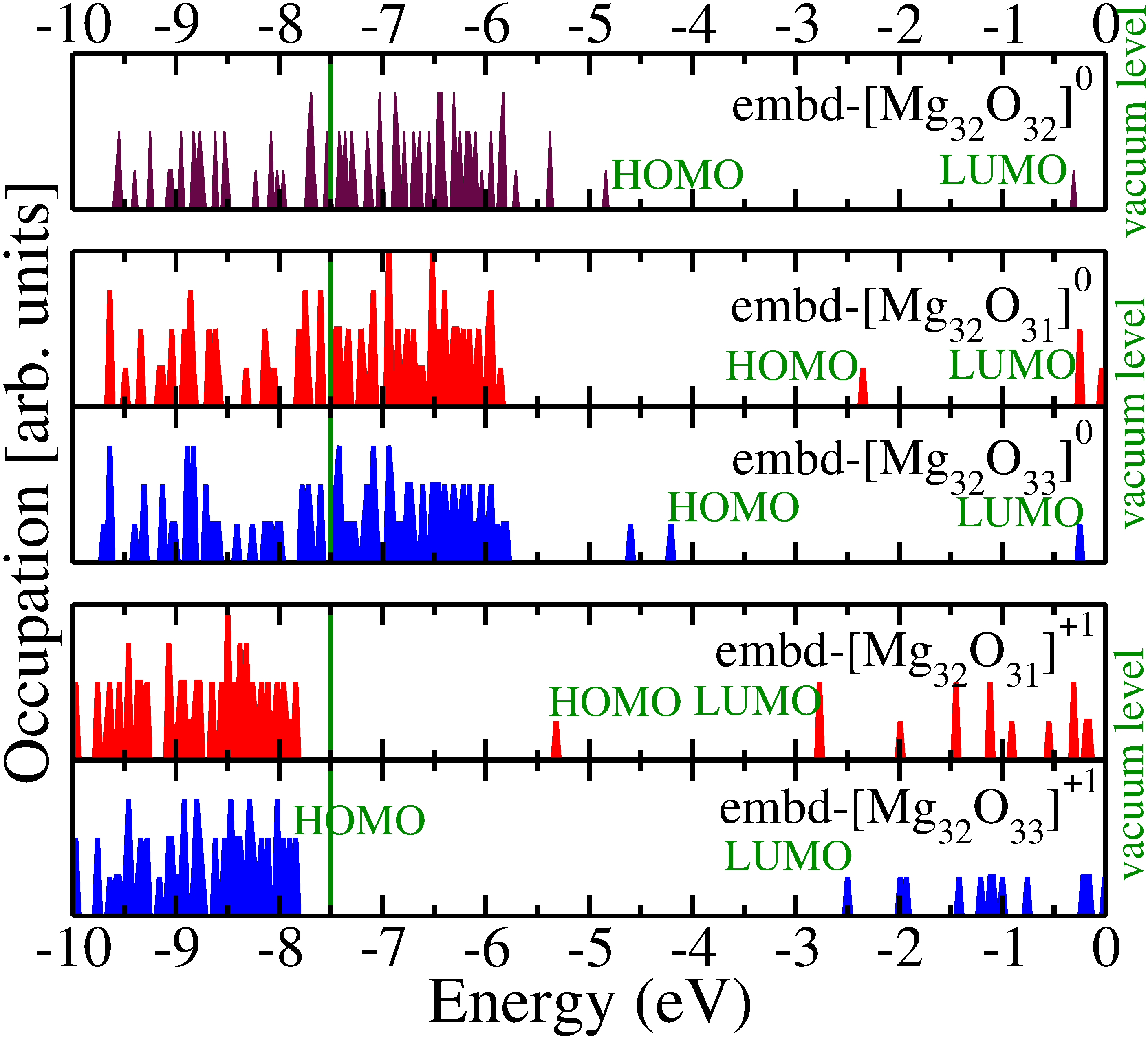}
\caption{(color online). Kohn-Sham spectrum of neutral embedded Mg$_{32}$O$_{32+x}$ cluster, with $x = -1$ (vacancy) and $x =1$ (O ad-species).}
\label{fig4}
\end{figure} 
In order to understand the extra stability of O$_1$/O$_2$-ad-species over O-vacancies at the corners of MgO surface, we analyze the density of states of embedded Mg$_{32}$O$_{32+x}$ clusters (Fig.~\ref{fig4}). The green line at -7.5~eV is the bulk Fermi level. The overall defect formation energy can be decomposed in two parts: (i) breaking/making the chemical bonds, and (ii) trapping the charge carriers. For the vacancy, the first part amounts to 5.08~eV. If we estimate the charging energy as the energy difference between the HOMO and the Fermi level, the vacancy can gain 6.48~eV by trapping two holes (see Fig.~\ref{fig4}), resulting in the overall estimated formation energy of -1.4~eV. While for the O-ad-species the estimated energy gain due to hole trapping is much smaller (2.09~eV), the neutral defect has also a much lower formation energy (0.21~eV), resulting in the overall estimated formation energy of $-2.09$~eV. Thus, the effect of the defect site coordination on the interplay of bond breaking/making and charge carrier trapping is responsible for the increased stability of O-ad-species. 

To confirm the existence of O$_1$ adsorbed species, (vibrational-spectroscopy) experiments on rugged MgO surfaces in a controlled O$_2$ atmosphere could look for the vibrational signature of the O-O bond stretching of adsorbed O$_1$ with a single trapped hole. We have calculated this frequency using PBE and HSE06 functionals, and the results are 658~cm$^{-1}$ and 728~cm$^{-1}$, respectively.~\footnote{For reference, PBE (HSE06) frequencies for gas-phase O$_2^-$ and O$_2^{2-}$ are 1113 (1115)~cm$^{-1}$ and 645 (669)~cm$^{-1}$, respectively. The experimental value for the gas-phase O$_2^-$ vibrational frequency is 1108~cm$^{-1}$~\cite{ervin2003only}, and for O$_2^{2-}$ in various dioxygen-metal complexes it is in the range 790-932~cm$^{-1}$~\cite{vaska1976dioxygen}.}

In conclusion, we find that changing a surface site coordination can change and even re-order the relative stability of surface sites in terms of oxidation or reduction. The applied methodology and the resulting knowledge can be used to design functional materials whose applications rely on surface chemistry.\\
\indent We are grateful to Matthias Scheffler for support and fruitful discussions. We thank the DFG cluster of excellence ``Unifying Concepts in Catalysis" (UniCat) for financial support. SB acknowledges the computing facility of HPC, IIT Delhi and Rechenzentrum Garching (RZG) of the Max-Planck Gesellschaft. SB acknowledges the financial support from YSS-SERB research grant, DST, India (grant no. YSS/2015/001209).

%


\begin{thebibliography}{28}%
\makeatletter
\providecommand \@ifxundefined [1]{%
 \@ifx{#1\undefined}
}%
\providecommand \@ifnum [1]{%
 \ifnum #1\expandafter \@firstoftwo
 \else \expandafter \@secondoftwo
 \fi
}%
\providecommand \@ifx [1]{%
 \ifx #1\expandafter \@firstoftwo
 \else \expandafter \@secondoftwo
 \fi
}%
\providecommand \natexlab [1]{#1}%
\providecommand \enquote  [1]{``#1''}%
\providecommand \bibnamefont  [1]{#1}%
\providecommand \bibfnamefont [1]{#1}%
\providecommand \citenamefont [1]{#1}%
\providecommand \href@noop [0]{\@secondoftwo}%
\providecommand \href [0]{\begingroup \@sanitize@url \@href}%
\providecommand \@href[1]{\@@startlink{#1}\@@href}%
\providecommand \@@href[1]{\endgroup#1\@@endlink}%
\providecommand \@sanitize@url [0]{\catcode `\\12\catcode `\$12\catcode
  `\&12\catcode `\#12\catcode `\^12\catcode `\_12\catcode `\%12\relax}%
\providecommand \@@startlink[1]{}%
\providecommand \@@endlink[0]{}%
\providecommand \url  [0]{\begingroup\@sanitize@url \@url }%
\providecommand \@url [1]{\endgroup\@href {#1}{\urlprefix }}%
\providecommand \urlprefix  [0]{URL }%
\providecommand \Eprint [0]{\href }%
\providecommand \doibase [0]{http://dx.doi.org/}%
\providecommand \selectlanguage [0]{\@gobble}%
\providecommand \bibinfo  [0]{\@secondoftwo}%
\providecommand \bibfield  [0]{\@secondoftwo}%
\providecommand \translation [1]{[#1]}%
\providecommand \BibitemOpen [0]{}%
\providecommand \bibitemStop [0]{}%
\providecommand \bibitemNoStop [0]{.\EOS\space}%
\providecommand \EOS [0]{\spacefactor3000\relax}%
\providecommand \BibitemShut  [1]{\csname bibitem#1\endcsname}%
\let\auto@bib@innerbib\@empty
\bibitem [{\citenamefont {Yan}\ \emph {et~al.}(2005)\citenamefont {Yan},
  \citenamefont {Chinta}, \citenamefont {Mohamed}, \citenamefont {Fackler},\
  and\ \citenamefont {Goodman}}]{Zhen2005}%
  \BibitemOpen
  \bibfield  {author} {\bibinfo {author} {\bibfnamefont {Z.}~\bibnamefont
  {Yan}}, \bibinfo {author} {\bibfnamefont {S.}~\bibnamefont {Chinta}},
  \bibinfo {author} {\bibfnamefont {A.~A.}\ \bibnamefont {Mohamed}}, \bibinfo
  {author} {\bibfnamefont {J.~P.}\ \bibnamefont {Fackler}}, \ and\ \bibinfo
  {author} {\bibfnamefont {D.~W.}\ \bibnamefont {Goodman}},\ }\href {\doibase
  10.1021/ja043652m} {\bibfield  {journal} {\bibinfo  {journal} {Journal of the
  American Chemical Society}\ }\textbf {\bibinfo {volume} {127}},\ \bibinfo
  {pages} {1604} (\bibinfo {year} {2005})},\ \bibinfo {note} {pMID: 15700971},\
  \Eprint {http://arxiv.org/abs/http://dx.doi.org/10.1021/ja043652m}
  {http://dx.doi.org/10.1021/ja043652m} \BibitemShut {NoStop}%
\bibitem [{\citenamefont {Balint}\ and\ \citenamefont
  {Aika}(2001)}]{Balint2001}%
  \BibitemOpen
  \bibfield  {author} {\bibinfo {author} {\bibfnamefont {I.}~\bibnamefont
  {Balint}}\ and\ \bibinfo {author} {\bibfnamefont {K.-i.}\ \bibnamefont
  {Aika}},\ }\href@noop {} {\bibfield  {journal} {\bibinfo  {journal} {Applied
  surface science}\ }\textbf {\bibinfo {volume} {173}},\ \bibinfo {pages} {296}
  (\bibinfo {year} {2001})}\BibitemShut {NoStop}%
\bibitem [{\citenamefont {Richter}\ \emph {et~al.}(2013)\citenamefont
  {Richter}, \citenamefont {Sicolo}, \citenamefont {Levchenko}, \citenamefont
  {Sauer},\ and\ \citenamefont {Scheffler}}]{Richter2013}%
  \BibitemOpen
  \bibfield  {author} {\bibinfo {author} {\bibfnamefont {N.~A.}\ \bibnamefont
  {Richter}}, \bibinfo {author} {\bibfnamefont {S.}~\bibnamefont {Sicolo}},
  \bibinfo {author} {\bibfnamefont {S.~V.}\ \bibnamefont {Levchenko}}, \bibinfo
  {author} {\bibfnamefont {J.}~\bibnamefont {Sauer}}, \ and\ \bibinfo {author}
  {\bibfnamefont {M.}~\bibnamefont {Scheffler}},\ }\href@noop {} {\bibfield
  {journal} {\bibinfo  {journal} {Phys. Rev. Lett.}\ }\textbf {\bibinfo
  {volume} {111}},\ \bibinfo {pages} {045502} (\bibinfo {year}
  {2013})}\BibitemShut {NoStop}%
\bibitem [{\citenamefont {Sushko}\ \emph {et~al.}(2000)\citenamefont {Sushko},
  \citenamefont {Shluger},\ and\ \citenamefont {Catlow}}]{Sushko2000153}%
  \BibitemOpen
  \bibfield  {author} {\bibinfo {author} {\bibfnamefont {P.~V.}\ \bibnamefont
  {Sushko}}, \bibinfo {author} {\bibfnamefont {A.~L.}\ \bibnamefont {Shluger}},
  \ and\ \bibinfo {author} {\bibfnamefont {C.~A.}\ \bibnamefont {Catlow}},\
  }\href {\doibase http://dx.doi.org/10.1016/S0039-6028(00)00290-9} {\bibfield
  {journal} {\bibinfo  {journal} {Surface Science}\ }\textbf {\bibinfo {volume}
  {450}},\ \bibinfo {pages} {153 } (\bibinfo {year} {2000})}\BibitemShut
  {NoStop}%
\bibitem [{\citenamefont {Sicolo}\ and\ \citenamefont
  {Sauer}(2013)}]{Sicolo2013}%
  \BibitemOpen
  \bibfield  {author} {\bibinfo {author} {\bibfnamefont {S.}~\bibnamefont
  {Sicolo}}\ and\ \bibinfo {author} {\bibfnamefont {J.}~\bibnamefont {Sauer}},\
  }\href@noop {} {\bibfield  {journal} {\bibinfo  {journal} {J. Phys. Chem. C}\
  }\textbf {\bibinfo {volume} {117}},\ \bibinfo {pages} {8365} (\bibinfo {year}
  {2013})}\BibitemShut {NoStop}%
\bibitem [{\citenamefont {Bhattacharya}\ \emph {et~al.}(2013)\citenamefont
  {Bhattacharya}, \citenamefont {Levchenko}, \citenamefont {Ghiringhelli},\
  and\ \citenamefont {Scheffler}}]{Bhattacharya2013}%
  \BibitemOpen
  \bibfield  {author} {\bibinfo {author} {\bibfnamefont {S.}~\bibnamefont
  {Bhattacharya}}, \bibinfo {author} {\bibfnamefont {S.~V.}\ \bibnamefont
  {Levchenko}}, \bibinfo {author} {\bibfnamefont {L.~M.}\ \bibnamefont
  {Ghiringhelli}}, \ and\ \bibinfo {author} {\bibfnamefont {M.}~\bibnamefont
  {Scheffler}},\ }\href@noop {} {\bibfield  {journal} {\bibinfo  {journal}
  {Phys. Rev. Lett.}\ }\textbf {\bibinfo {volume} {111}},\ \bibinfo {pages}
  {135501} (\bibinfo {year} {2013})}\BibitemShut {NoStop}%
\bibitem [{Note1()}]{Note1}%
  \BibitemOpen
  \bibinfo {note} {A semiconducting $p$-doped MgO can be produced by, e.g.,
  doping with Li~\cite {Chen1980,tardio2002p}. Even without intentional doping,
  nominally pure MgO becomes a $p$-type semiconductor at temperatures above
  800~K~\cite {Freund1993}, which is comparable or even below the typical
  temperatures for some of the catalytic applications of MgO.}\BibitemShut
  {Stop}%
\bibitem [{\citenamefont {Scheffler}\ and\ \citenamefont
  {Weinert}(1986)}]{Scheffler1986}%
  \BibitemOpen
  \bibfield  {author} {\bibinfo {author} {\bibfnamefont {M.}~\bibnamefont
  {Scheffler}}\ and\ \bibinfo {author} {\bibfnamefont {C.}~\bibnamefont
  {Weinert}},\ }in\ \href@noop {} {\emph {\bibinfo {booktitle} {Defects in
  Semiconductors}}},\ \bibinfo {editor} {edited by\ \bibinfo {editor}
  {\bibfnamefont {H.~J.~v.}\ \bibnamefont {Bardeleben}}}\ (\bibinfo
  {publisher} {Trans. Tech. Publ. Ltd, Switzerland},\ \bibinfo {year} {1986})\
  pp.\ \bibinfo {pages} {25--30}\BibitemShut {NoStop}%
\bibitem [{\citenamefont {Bernstein}\ \emph {et~al.}(2009)\citenamefont
  {Bernstein}, \citenamefont {Kermode},\ and\ \citenamefont
  {Cs\'{a}nyi}}]{Bernstein2009}%
  \BibitemOpen
  \bibfield  {author} {\bibinfo {author} {\bibfnamefont {N.}~\bibnamefont
  {Bernstein}}, \bibinfo {author} {\bibfnamefont {J.~R.}\ \bibnamefont
  {Kermode}}, \ and\ \bibinfo {author} {\bibfnamefont {G.}~\bibnamefont
  {Cs\'{a}nyi}},\ }\href@noop {} {\bibfield  {journal} {\bibinfo  {journal}
  {Rep. Prog. Phys.}\ }\textbf {\bibinfo {volume} {72}},\ \bibinfo {pages}
  {026501} (\bibinfo {year} {2009})}\BibitemShut {NoStop}%
\bibitem [{\citenamefont {Kleinman}\ and\ \citenamefont
  {Bylander}(1982)}]{KleinmanBylander1982}%
  \BibitemOpen
  \bibfield  {author} {\bibinfo {author} {\bibfnamefont {L.}~\bibnamefont
  {Kleinman}}\ and\ \bibinfo {author} {\bibfnamefont {D.~M.}\ \bibnamefont
  {Bylander}},\ }\href@noop {} {\bibfield  {journal} {\bibinfo  {journal}
  {Phys. Rev. Lett.}\ }\textbf {\bibinfo {volume} {48}},\ \bibinfo {pages}
  {1425} (\bibinfo {year} {1982})}\BibitemShut {NoStop}%
\bibitem [{\citenamefont {Blum}\ \emph {et~al.}(2009)\citenamefont {Blum},
  \citenamefont {Gehrke}, \citenamefont {Hanke}, \citenamefont {Havu},
  \citenamefont {Havu}, \citenamefont {Ren}, \citenamefont {Reuter},\ and\
  \citenamefont {Scheffler}}]{Blum2009}%
  \BibitemOpen
  \bibfield  {author} {\bibinfo {author} {\bibfnamefont {V.}~\bibnamefont
  {Blum}}, \bibinfo {author} {\bibfnamefont {R.}~\bibnamefont {Gehrke}},
  \bibinfo {author} {\bibfnamefont {F.}~\bibnamefont {Hanke}}, \bibinfo
  {author} {\bibfnamefont {P.}~\bibnamefont {Havu}}, \bibinfo {author}
  {\bibfnamefont {V.}~\bibnamefont {Havu}}, \bibinfo {author} {\bibfnamefont
  {X.}~\bibnamefont {Ren}}, \bibinfo {author} {\bibfnamefont {K.}~\bibnamefont
  {Reuter}}, \ and\ \bibinfo {author} {\bibfnamefont {M.}~\bibnamefont
  {Scheffler}},\ }\href@noop {} {\bibfield  {journal} {\bibinfo  {journal}
  {Comput. Phys. Commun.}\ }\textbf {\bibinfo {volume} {180}},\ \bibinfo
  {pages} {2175} (\bibinfo {year} {2009})}\BibitemShut {NoStop}%
\bibitem [{\citenamefont {Berger}\ \emph {et~al.}(2014)\citenamefont {Berger},
  \citenamefont {Logsdail}, \citenamefont {Oberhofer}, \citenamefont {Farrow},
  \citenamefont {Catlow}, \citenamefont {Sherwood}, \citenamefont {Sokol},
  \citenamefont {Blum},\ and\ \citenamefont {Reuter}}]{Berger2014}%
  \BibitemOpen
  \bibfield  {author} {\bibinfo {author} {\bibfnamefont {D.}~\bibnamefont
  {Berger}}, \bibinfo {author} {\bibfnamefont {A.~J.}\ \bibnamefont
  {Logsdail}}, \bibinfo {author} {\bibfnamefont {H.}~\bibnamefont {Oberhofer}},
  \bibinfo {author} {\bibfnamefont {M.~R.}\ \bibnamefont {Farrow}}, \bibinfo
  {author} {\bibfnamefont {C.~R.~A.}\ \bibnamefont {Catlow}}, \bibinfo {author}
  {\bibfnamefont {P.}~\bibnamefont {Sherwood}}, \bibinfo {author}
  {\bibfnamefont {A.~A.}\ \bibnamefont {Sokol}}, \bibinfo {author}
  {\bibfnamefont {V.}~\bibnamefont {Blum}}, \ and\ \bibinfo {author}
  {\bibfnamefont {K.}~\bibnamefont {Reuter}},\ }\href {\doibase
  10.1063/1.4885816} {\bibfield  {journal} {\bibinfo  {journal} {The Journal of
  Chemical Physics}\ }\textbf {\bibinfo {volume} {141}},\ \bibinfo {pages}
  {024105} (\bibinfo {year} {2014})},\ \Eprint
  {http://arxiv.org/abs/http://dx.doi.org/10.1063/1.4885816}
  {http://dx.doi.org/10.1063/1.4885816} \BibitemShut {NoStop}%
\bibitem [{\citenamefont {Berger}\ \emph {et~al.}(2015)\citenamefont {Berger},
  \citenamefont {Oberhofer},\ and\ \citenamefont {Reuter}}]{Berger2015}%
  \BibitemOpen
  \bibfield  {author} {\bibinfo {author} {\bibfnamefont {D.}~\bibnamefont
  {Berger}}, \bibinfo {author} {\bibfnamefont {H.}~\bibnamefont {Oberhofer}}, \
  and\ \bibinfo {author} {\bibfnamefont {K.}~\bibnamefont {Reuter}},\ }\href
  {\doibase 10.1103/PhysRevB.92.075308} {\bibfield  {journal} {\bibinfo
  {journal} {Phys. Rev. B}\ }\textbf {\bibinfo {volume} {92}},\ \bibinfo
  {pages} {075308} (\bibinfo {year} {2015})}\BibitemShut {NoStop}%
\bibitem [{\citenamefont {Sherwood}\ \emph {et~al.}(2003)\citenamefont
  {Sherwood}, \citenamefont {de~Vries}, \citenamefont {Guest}, \citenamefont
  {Schreckenbach}, \citenamefont {Catlow}, \citenamefont {French},
  \citenamefont {Sokol}, \citenamefont {Bromley}, \citenamefont {Thiel},
  \citenamefont {Turner}, \citenamefont {Billeter}, \citenamefont {Terstegen},
  \citenamefont {Thiel}, \citenamefont {Kendrick}, \citenamefont {Rogers},
  \citenamefont {Casci}, \citenamefont {Watson}, \citenamefont {King},
  \citenamefont {Karlsen}, \citenamefont {Sj\"{o}voll}, \citenamefont {Fahmi},
  \citenamefont {Sch\"{a}fer},\ and\ \citenamefont {Lennartz}}]{Sherwood2003}%
  \BibitemOpen
  \bibfield  {author} {\bibinfo {author} {\bibfnamefont {P.}~\bibnamefont
  {Sherwood}}, \bibinfo {author} {\bibfnamefont {A.~H.}\ \bibnamefont
  {de~Vries}}, \bibinfo {author} {\bibfnamefont {M.~F.}\ \bibnamefont {Guest}},
  \bibinfo {author} {\bibfnamefont {G.}~\bibnamefont {Schreckenbach}}, \bibinfo
  {author} {\bibfnamefont {C.~R.~A.}\ \bibnamefont {Catlow}}, \bibinfo {author}
  {\bibfnamefont {S.~A.}\ \bibnamefont {French}}, \bibinfo {author}
  {\bibfnamefont {A.~A.}\ \bibnamefont {Sokol}}, \bibinfo {author}
  {\bibfnamefont {S.~T.}\ \bibnamefont {Bromley}}, \bibinfo {author}
  {\bibfnamefont {W.}~\bibnamefont {Thiel}}, \bibinfo {author} {\bibfnamefont
  {A.~J.}\ \bibnamefont {Turner}}, \bibinfo {author} {\bibfnamefont
  {S.}~\bibnamefont {Billeter}}, \bibinfo {author} {\bibfnamefont
  {F.}~\bibnamefont {Terstegen}}, \bibinfo {author} {\bibfnamefont
  {S.}~\bibnamefont {Thiel}}, \bibinfo {author} {\bibfnamefont
  {J.}~\bibnamefont {Kendrick}}, \bibinfo {author} {\bibfnamefont {S.~C.}\
  \bibnamefont {Rogers}}, \bibinfo {author} {\bibfnamefont {J.}~\bibnamefont
  {Casci}}, \bibinfo {author} {\bibfnamefont {M.}~\bibnamefont {Watson}},
  \bibinfo {author} {\bibfnamefont {F.}~\bibnamefont {King}}, \bibinfo {author}
  {\bibfnamefont {E.}~\bibnamefont {Karlsen}}, \bibinfo {author} {\bibfnamefont
  {M.}~\bibnamefont {Sj\"{o}voll}}, \bibinfo {author} {\bibfnamefont
  {A.}~\bibnamefont {Fahmi}}, \bibinfo {author} {\bibfnamefont
  {A.}~\bibnamefont {Sch\"{a}fer}}, \ and\ \bibinfo {author} {\bibfnamefont
  {C.}~\bibnamefont {Lennartz}},\ }\href@noop {} {\bibfield  {journal}
  {\bibinfo  {journal} {J. Mol. Struct. (Theochem.)}\ }\textbf {\bibinfo
  {volume} {632}},\ \bibinfo {pages} {1} (\bibinfo {year} {2003})}\BibitemShut
  {NoStop}%
\bibitem [{\citenamefont {Sokol}\ \emph {et~al.}(2004)\citenamefont {Sokol},
  \citenamefont {Bromley}, \citenamefont {French}, \citenamefont {Catlow},\
  and\ \citenamefont {Sherwood}}]{Sokol2004}%
  \BibitemOpen
  \bibfield  {author} {\bibinfo {author} {\bibfnamefont {A.~A.}\ \bibnamefont
  {Sokol}}, \bibinfo {author} {\bibfnamefont {S.~T.}\ \bibnamefont {Bromley}},
  \bibinfo {author} {\bibfnamefont {S.~A.}\ \bibnamefont {French}}, \bibinfo
  {author} {\bibfnamefont {C.~R.~A.}\ \bibnamefont {Catlow}}, \ and\ \bibinfo
  {author} {\bibfnamefont {P.}~\bibnamefont {Sherwood}},\ }\href@noop {}
  {\bibfield  {journal} {\bibinfo  {journal} {Int. J. Quantum Chem}\ }\textbf
  {\bibinfo {volume} {99}},\ \bibinfo {pages} {695} (\bibinfo {year}
  {2004})}\BibitemShut {NoStop}%
\bibitem [{\citenamefont {Gale}(1997)}]{GULP}%
  \BibitemOpen
  \bibfield  {author} {\bibinfo {author} {\bibfnamefont {J.~D.}\ \bibnamefont
  {Gale}},\ }\href@noop {} {\bibfield  {journal} {\bibinfo  {journal} {J. Chem.
  Soc.{,} Faraday Trans.}\ }\textbf {\bibinfo {volume} {93}},\ \bibinfo {pages}
  {629} (\bibinfo {year} {1997})}\BibitemShut {NoStop}%
\bibitem [{\citenamefont {Perdew}\ \emph {et~al.}(1996)\citenamefont {Perdew},
  \citenamefont {Burke},\ and\ \citenamefont {Ernzerhof}}]{Perdew1996}%
  \BibitemOpen
  \bibfield  {author} {\bibinfo {author} {\bibfnamefont {J.}~\bibnamefont
  {Perdew}}, \bibinfo {author} {\bibfnamefont {K.}~\bibnamefont {Burke}}, \
  and\ \bibinfo {author} {\bibfnamefont {M.}~\bibnamefont {Ernzerhof}},\
  }\href@noop {} {\bibfield  {journal} {\bibinfo  {journal} {Phys. Rev. Lett.}\
  }\textbf {\bibinfo {volume} {77}},\ \bibinfo {pages} {3865} (\bibinfo {year}
  {1996})}\BibitemShut {NoStop}%
\bibitem [{\citenamefont {Perdew}\ \emph {et~al.}(1997)\citenamefont {Perdew},
  \citenamefont {Burke},\ and\ \citenamefont {Ernzerhof}}]{Perdew1997}%
  \BibitemOpen
  \bibfield  {author} {\bibinfo {author} {\bibfnamefont {J.}~\bibnamefont
  {Perdew}}, \bibinfo {author} {\bibfnamefont {K.}~\bibnamefont {Burke}}, \
  and\ \bibinfo {author} {\bibfnamefont {M.}~\bibnamefont {Ernzerhof}},\
  }\href@noop {} {\bibfield  {journal} {\bibinfo  {journal} {Phys. Rev. Lett.}\
  }\textbf {\bibinfo {volume} {78}},\ \bibinfo {pages} {1396} (\bibinfo {year}
  {1997})}\BibitemShut {NoStop}%
\bibitem [{\citenamefont {Heyd}\ \emph {et~al.}(2006)\citenamefont {Heyd},
  \citenamefont {Scuseria},\ and\ \citenamefont {Ernzerhof}}]{Heyd2006}%
  \BibitemOpen
  \bibfield  {author} {\bibinfo {author} {\bibfnamefont {J.}~\bibnamefont
  {Heyd}}, \bibinfo {author} {\bibfnamefont {G.~E.}\ \bibnamefont {Scuseria}},
  \ and\ \bibinfo {author} {\bibfnamefont {M.}~\bibnamefont {Ernzerhof}},\
  }\href@noop {} {\bibfield  {journal} {\bibinfo  {journal} {J. Chem. Phys.}\
  }\textbf {\bibinfo {volume} {124}},\ \bibinfo {pages} {219906} (\bibinfo
  {year} {2006})}\BibitemShut {NoStop}%
\bibitem [{\citenamefont {Tkatchenko}\ and\ \citenamefont
  {Scheffler}(2009)}]{Tkatchenko2009}%
  \BibitemOpen
  \bibfield  {author} {\bibinfo {author} {\bibfnamefont {A.}~\bibnamefont
  {Tkatchenko}}\ and\ \bibinfo {author} {\bibfnamefont {M.}~\bibnamefont
  {Scheffler}},\ }\href@noop {} {\bibfield  {journal} {\bibinfo  {journal}
  {Phys. Rev. Lett.}\ }\textbf {\bibinfo {volume} {102}},\ \bibinfo {pages}
  {073005} (\bibinfo {year} {2009})}\BibitemShut {NoStop}%
\bibitem [{\citenamefont {Bhattacharya}\ \emph {et~al.}(2014)\citenamefont
  {Bhattacharya}, \citenamefont {Levchenko}, \citenamefont {Ghiringhelli},\
  and\ \citenamefont {Scheffler}}]{Bhattacharya2014}%
  \BibitemOpen
  \bibfield  {author} {\bibinfo {author} {\bibfnamefont {S.}~\bibnamefont
  {Bhattacharya}}, \bibinfo {author} {\bibfnamefont {S.~V.}\ \bibnamefont
  {Levchenko}}, \bibinfo {author} {\bibfnamefont {L.~M.}\ \bibnamefont
  {Ghiringhelli}}, \ and\ \bibinfo {author} {\bibfnamefont {M.}~\bibnamefont
  {Scheffler}},\ }\href {http://stacks.iop.org/1367-2630/16/i=12/a=123016}
  {\bibfield  {journal} {\bibinfo  {journal} {New Journal of Physics}\ }\textbf
  {\bibinfo {volume} {16}},\ \bibinfo {pages} {123016} (\bibinfo {year}
  {2014})}\BibitemShut {NoStop}%
\bibitem [{\citenamefont {Plass}\ \emph {et~al.}(1998)\citenamefont {Plass},
  \citenamefont {Egan}, \citenamefont {Collazo-Davila}, \citenamefont {Grozea},
  \citenamefont {Landree}, \citenamefont {Marks},\ and\ \citenamefont
  {Gajdardziska-Josifovska}}]{Richard_PRL1998}%
  \BibitemOpen
  \bibfield  {author} {\bibinfo {author} {\bibfnamefont {R.}~\bibnamefont
  {Plass}}, \bibinfo {author} {\bibfnamefont {K.}~\bibnamefont {Egan}},
  \bibinfo {author} {\bibfnamefont {C.}~\bibnamefont {Collazo-Davila}},
  \bibinfo {author} {\bibfnamefont {D.}~\bibnamefont {Grozea}}, \bibinfo
  {author} {\bibfnamefont {E.}~\bibnamefont {Landree}}, \bibinfo {author}
  {\bibfnamefont {L.~D.}\ \bibnamefont {Marks}}, \ and\ \bibinfo {author}
  {\bibfnamefont {M.}~\bibnamefont {Gajdardziska-Josifovska}},\ }\href
  {\doibase 10.1103/PhysRevLett.81.4891} {\bibfield  {journal} {\bibinfo
  {journal} {Phys. Rev. Lett.}\ }\textbf {\bibinfo {volume} {81}},\ \bibinfo
  {pages} {4891} (\bibinfo {year} {1998})}\BibitemShut {NoStop}%
\bibitem [{Note2()}]{Note2}%
  \BibitemOpen
  \bibinfo {note} {For reference, PBE (HSE06) frequencies for gas-phase O$_2^-$
  and O$_2^{2-}$ are 1113 (1115)~cm$^{-1}$ and 645 (669)~cm$^{-1}$,
  respectively. The experimental value for the gas-phase O$_2^-$ vibrational
  frequency is 1108~cm$^{-1}$~\cite {ervin2003only}, and for O$_2^{2-}$ in
  various dioxygen-metal complexes it is in the range 790-932~cm$^{-1}$~\cite
  {vaska1976dioxygen}.}\BibitemShut {Stop}%
\bibitem [{\citenamefont {Chen}\ \emph {et~al.}(1980)\citenamefont {Chen},
  \citenamefont {Boldu},\ and\ \citenamefont {Orera}}]{Chen1980}%
  \BibitemOpen
  \bibfield  {author} {\bibinfo {author} {\bibfnamefont {Y.}~\bibnamefont
  {Chen}}, \bibinfo {author} {\bibfnamefont {J.}~\bibnamefont {Boldu}}, \ and\
  \bibinfo {author} {\bibfnamefont {V.}~\bibnamefont {Orera}},\ }\href@noop {}
  {\bibfield  {journal} {\bibinfo  {journal} {Le Journal de Physique
  Colloques}\ }\textbf {\bibinfo {volume} {41}},\ \bibinfo {pages} {C6}
  (\bibinfo {year} {1980})}\BibitemShut {NoStop}%
\bibitem [{\citenamefont {Tard{\'\i}o}\ \emph {et~al.}(2002)\citenamefont
  {Tard{\'\i}o}, \citenamefont {Ram{\'\i}rez}, \citenamefont {Gonz{\'a}lez},\
  and\ \citenamefont {Chen}}]{tardio2002p}%
  \BibitemOpen
  \bibfield  {author} {\bibinfo {author} {\bibfnamefont {M.}~\bibnamefont
  {Tard{\'\i}o}}, \bibinfo {author} {\bibfnamefont {R.}~\bibnamefont
  {Ram{\'\i}rez}}, \bibinfo {author} {\bibfnamefont {R.}~\bibnamefont
  {Gonz{\'a}lez}}, \ and\ \bibinfo {author} {\bibfnamefont {Y.}~\bibnamefont
  {Chen}},\ }\href@noop {} {\bibfield  {journal} {\bibinfo  {journal} {Physical
  Review B}\ }\textbf {\bibinfo {volume} {66}},\ \bibinfo {pages} {134202}
  (\bibinfo {year} {2002})}\BibitemShut {NoStop}%
\bibitem [{\citenamefont {Freund}\ \emph {et~al.}(1993)\citenamefont {Freund},
  \citenamefont {Freund},\ and\ \citenamefont {Batllo}}]{Freund1993}%
  \BibitemOpen
  \bibfield  {author} {\bibinfo {author} {\bibfnamefont {F.}~\bibnamefont
  {Freund}}, \bibinfo {author} {\bibfnamefont {M.~M.}\ \bibnamefont {Freund}},
  \ and\ \bibinfo {author} {\bibfnamefont {F.}~\bibnamefont {Batllo}},\
  }\href@noop {} {\bibfield  {journal} {\bibinfo  {journal} {Journal of
  Geophysical Research: Solid Earth}\ }\textbf {\bibinfo {volume} {98}},\
  \bibinfo {pages} {22209} (\bibinfo {year} {1993})}\BibitemShut {NoStop}%
\bibitem [{\citenamefont {Ervin}\ \emph {et~al.}(2003)\citenamefont {Ervin},
  \citenamefont {Anusiewicz}, \citenamefont {Skurski}, \citenamefont {Simons},\
  and\ \citenamefont {Lineberger}}]{ervin2003only}%
  \BibitemOpen
  \bibfield  {author} {\bibinfo {author} {\bibfnamefont {K.~M.}\ \bibnamefont
  {Ervin}}, \bibinfo {author} {\bibfnamefont {I.}~\bibnamefont {Anusiewicz}},
  \bibinfo {author} {\bibfnamefont {P.}~\bibnamefont {Skurski}}, \bibinfo
  {author} {\bibfnamefont {J.}~\bibnamefont {Simons}}, \ and\ \bibinfo {author}
  {\bibfnamefont {W.~C.}\ \bibnamefont {Lineberger}},\ }\href@noop {}
  {\bibfield  {journal} {\bibinfo  {journal} {The Journal of Physical Chemistry
  A}\ }\textbf {\bibinfo {volume} {107}},\ \bibinfo {pages} {8521} (\bibinfo
  {year} {2003})}\BibitemShut {NoStop}%
\bibitem [{\citenamefont {Vaska}(1976)}]{vaska1976dioxygen}%
  \BibitemOpen
  \bibfield  {author} {\bibinfo {author} {\bibfnamefont {L.}~\bibnamefont
  {Vaska}},\ }\href@noop {} {\bibfield  {journal} {\bibinfo  {journal}
  {Accounts of Chemical Research}\ }\textbf {\bibinfo {volume} {9}},\ \bibinfo
  {pages} {175} (\bibinfo {year} {1976})}\BibitemShut {NoStop}%
\end{thebibliography}
\end{document}